\documentstyle[preprint,aps]{revtex}
\begin{document}


\draft


\title{
\begin{flushright}
{\rm gr-qc/9408020}
\end{flushright}
 On the stability of black hole event horizons}


\author{Bj\o rn Jensen
\footnote{Electronic address: BJensen@boson.uio.no}}

\address{Institute of Physics, University of Oslo,
P.O. Box 1048, N-0316 Blindern, Oslo 3, Norway}

\date{\today}

\bibliographystyle{unsrt}

\maketitle


\begin{abstract}
In this work we study a {\it gedanken} experiment
constructed
in order to test the cosmic censorship hypothesis
and the second law of black hole
thermo-dynamics. Matter with a negative
gravitating energy is imagined
added to a near extremal $U(1)$-charged static
black hole in Einstein-Maxwell theory. The
dynamics of a similar process is studied and the
thermo-dynamical properties of the resulting
black hole structure is discussed. A new mechanism which
stabilizes black hole event horizons is shown
to operate in such processes.
\end{abstract}





\section{Introduction}

In classical general relativity theory
strong curvature singularities, i.e. regions of
space-time
through which no geodesic can be extended,
are predicted to appear in space-time.
Many believe that such
singularities
are always hidden behind
event-horizons. A cosmic censor will always hide
 the singular region
in such a way as to make the space-time maximally
predictable,
it is believed. This is the essence of a collection
 of proposals
collectively known as the cosmic censor-hypothesis
(for a review see eg.\cite{Clarke}). However,
even though general relativity predicts the existence
 of singular
regions accompanied by event-horizons when certain
restrictions
on the properties of the
the matter content of the world is made a proof of
the stability of
black hole event horizons does not exist
(see eg.\cite{Joshi}). Even though a singular
region is hidden behind an event-horizon initially
it is
unclear whether the horizon can be removed or not
in the future by a physical process so that the
singular
region is exposed.

Classically no process will shrink the
area of a single black hole in an open universe provided
 that
the strong energy condition \cite{Hawking} holds. This
is connected with the one
way nature of the black hole horizon; nothing can
 escape from
a black hole classically. When quantum effects are taken
 into account it is well
 known that
black hole horizons are destabilized, i.e. the surface
 area
of the horizon will decrease due to the emission of
quantum particles. The decrease of the surface area
can also be looked upon as due to the absorption by
the black hole
of negative energy from the quantum vacuum.
However, all
computations
of the Hawking process relies on the validity of
the semi-classical
approximation to quantum gravity and at the last
stages of the
evaporation process this approximation
generally breaks down. Hence, what sorts of
structures that will remain after the Hawking evaporation
 process has
come to a halt, i.e. whether only radiation
remains or a naked singularity or a massive
remnant is left behind, is completely unclear.

One of the curious properties
of general relativity theory is that it
assigns an
{\it absolute} thermo-dynamic
entropy to black holes in terms of the area
of the black hole horizon. Hence, the cosmic
censor-ship
hypothesis is thus seen to be linked to well
known
classical thermo-dynamics. The one-way nature
of the
event horizon on the classical level is in this way
connected to the
statement that $d(Entropy)\geq 0$. The Hawking
evaporation process
violates this relation since it makes the horizon
 area shrink and hence the associated entropy decrease.
However a generalized second law can be formulated
in such a
way that the entropy loss can be found in the
resulting
entropy increase in the radiation field outside the
black hole.

In this work we study a {\it gedanken} experiment
 devised
with the violation of the cosmic censor-ship
and the second law of black hole
thermo-dynamics in mind.
We will
focus the attention to a near extremal Reisner-
Nordstr\"{o}m
black hole (i.e. black hole charge $\sim$ black
hole metric mass) where thermal quantum effects are
highly suppressed.
We will
be concerned with the situation when macroscopic
entities carrying negative
gravitating energy (in violation of the strong
energy condition) are thrown into
 the extremal hole.
Naively one should expect that the mass of the resulting
structure could
become less than its charge
(in gravitational units $c=\hbar =k_B=G=1$).
When this happens the structure will display a
naked singularity.

The structure of this paper is as follows. In the next section we
 present
the {\it gedanken} experiment. In section III we
discuss some properties of negative mass carrying objects
 which generate a
spherical symmetric geometry and the actual innfall of
such an object into a black hole.
In section IV we study the consequences of this innfall in
relation to the cosmic censor-ship hypothesis and the
second law. We conclude with a section where we
summarize our findings.

\section{A gedanken experiment}

Consider a near extremal Reisner-Nordstr\"{o}m black hole, i.e.
an electrically
 $Q$ charged and metric mass $M$ carrying
static black hole in Einstein-Maxwell theory where the equality
 in
\begin{equation}
Q^2\leq M^2
\end{equation}
is nearly attained. In a recent work \cite{Becken} an attempt was
 made
to break this relation
by adding another $U(1)$ kind of charge $q$ to the black hole
in such a way that $q^2+Q^2>M^2$. Different kinds of processes
 where studied with both classical
and quantum mechanical $q$-charge carrying entities.
 It was concluded
that $q^2+Q^2\leq M^2$ probably
always holds in such processes due both to classical
effects and due to the charge
screening effect of the quantum-vacuum. In this work
another way to try to break the
relation eq.(1) is investigated. We will not try to add more
 charge of some kind but to try to
reduce to mass term in eq.(1). In the Hawking process this is
exactly
what is done but
when the black hole parameters reach a region of the parameter
space where
equility in eq.(1) is about to be attained the Hawking process
shuts down, i.e.
the temperature drops to zero. Hence, in this case the Hawking
process obeys the
cosmic censorship hypotesis. This is not necessarily a general
 feature of the Hawking radiation process since
it is known that electrically charged black holes in the low energy
 limit of the
heterotic string theory may display a naked singularity when it
radiates down to
extremality \cite{Holzhey}.

One important principle often assumed in the classical singularity
theorems in
the general theory of relativity is that the strong energy
condition is to
be satisfied \cite{Hawking}, i.e. the gravitating mass density $\rho_G$
satisfies
\begin{equation}
\rho_G\equiv -T^{\hat{t}}\, _{\hat{t}} +T^{\hat{i}}\,_{\hat{i}}\geq 0
\end{equation}
relative to an arbitrary tetrad (or null-frame).
Here $T$ denotes the
matter energy-momentum tensor
$\hat{t}$ the timelike indice and $\hat{i}\neq\hat{t}$.
Two initially parallel geodesics will always converge when
 passing
trough matter that obey this ``timelike convergence
condition'' and as a consequence
curvature
singularities must appear in the theory. However the strong
 energy condition
is probably a to restrictive condition to put on matter in
order to classify
a specific chunk of matter as physically acceptable. Probably
 the larger part of the
total energy density of the universe is contributed
 from the quantum vacuum
where the pressures in the three space-like dimensions
 equals minus a
small positive vacuum energy density $\rho$. It follows that the
gravitating mass density
contribution of the vacuum  is $\rho_G=-2\rho$ in violation of eq.(2).

In our {\it gedanken} experiment we will specifically imagine
that an object of finite spatial extend and with
an energy-momentum tensor with the above characteristics
of the vacuum is lowered into an extremal Reisner-Nordstr\"{o}m
black hole. Such objects was probably produced copiously in
phase-transitions in the early universe in the form of
topological defects \cite{Vilenkin1}. We will briefly review the most
important properties of some of these objects in the
next section. We expect that adding such ultra-relativistic
matter to a spesific system will imply that the resulting gravitating
mass of the system will be lower than the mass of the system before the
negative energy where added.
Is it possible to add sufficient
negative mass such that the
relation in eq.(1) is violated ?
\footnote{In \cite{Ford2,Ford3} energy which breaks the weak energy
condition was imagined injected into a near extremal Reisner-Nordstr\"{o}m
black hole. It was concluded that quantum field theory prevents an
unambiguous violation of cosmic censorship in such a process
when the averaged weak energy condition is assumed to hold.}

\section{Objects with negative mass}

In pure Einstein theory it has been proved that the
total
energy (the ADM mass) carried by an isolated system,
i.e. one that
generates an
asymptotic Minkowski geometry, is non-negative \cite{Schoen}.
Objects with a negative mass is therefore expected to
generate a curved asymptotic structure. Simple spherically symmetric
entities with this property is topological defects in
field theories where global symmetries are broken.
Examples of such defects are the global
monopoles arising in a $\phi^4$-theory with an internal
$O(3)$ global symmetry \cite{Barriola} and skyrmions which arise in
a non-linear $\sigma$-model with a $SO(4)$ symmetry
(see eg.\cite{Lee}).
In the asymptotic region $r\rightarrow\infty$ far from the
core-region of these defects
we find that
the geometric structures become
\begin{equation}
ds^2=-dt^2+dr^2+\alpha^2 r^2(d\theta^2+\sin^2\theta d\phi^2)
\end{equation}
where $\theta$ and $\phi$ is the usual polar angles.
It is clear
that this metric represents a curved space since
it can not be transformed to the Minkowski metric
in polar coordinates. The Ricci curvature scalar $R$ is
readily computed
\begin{equation}
R=\frac{(1-\alpha )(1+\alpha )}{\alpha^2 r^2}\, .
\end{equation}
{}From this expression it seems that the intrinsic
curvature vanishes in the limit
$r\rightarrow\infty$ \cite{Lee}. However this
is not correct since the expression for $R$ refers to
the behavior of the scalar curvature relative to a
specified point. Any other point in the assymptotic region
will also locally be endowed with the metric eq.(3)
and an $R$ on the form eq.(4) will be measured. In fact
a $\alpha\neq 1$ will induce
an effective curvature to the universe \cite{Jensen3}.

As was mentioned above since
the assymptotic regions of the above sources are curved
it follows that the total gravitating mass of these
objects can be expected to be negative.
Indeed, in \cite{Harari} a model of the interior of a global
monopole
where presented which was made up
of a part of the de-Sitter manifold.
This interior region could be matched smoothly to a
spherical symmetric geometry on the Schwartzshild form
but only with a deficit cone in the geometry. This
of-course gives rise to a negative mass object since the
mass density $\rho >0$ combined with local Lorentz invariance
of the source in the interior region implies
a negative gravitating mass as explained above
\cite{Jensen3}. It is worth noting that the ``vacuum'' outside
the global monopole can be modeled by an energy-momentum
tensor with vanishing angular components but where
$T^t\, _t=T^r\, _r\neq 0$ \cite{Barriola}. This Lorentz invariance in the
radial direction is directly linked to the
existence of the deficit cone. The
gravitational mass density $\rho_G$ of the field outside
the core-region vanishes due to
this invariance property.

The study of the innfall of a particle with a deficit cone
into a black hole is
very complicated. The process can not be treated
in a consistent manner in the test-particle
approximation since we never can neglect the effects produced
by the deficit cone. We will therefore consider a related
scenario where a spherically symmetric domain wall
$\Sigma$ with a black hole
in the center and with a deficit cone in the exterior region
collapses into the black hole. The very definition of a domain wall
implies that such structures will carry a negative gravitational energy.
Even though these processes
may seem unrelated the net effects produced in the two
scenarios on the hole is expected to be the
same since the resulting black hole will not carry any information
on the details of the innfall process.

We will first
study the properties of a static spherically
symmetric domain wall with and without a black hole
in the center of the spherical geometry.
To this end
we will utilize the Israel formalism.
The energy-stress
tensor $S^i\, _j$ in $\Sigma$ is determined by Israel's
equations \cite{Israel}
\begin{equation}
\gamma^i\, _j-\delta^i\, _j{\mbox Tr}\gamma_{ij} =-8\pi GS^i\, _j
\end{equation}
where $\gamma^i\, _j=\Delta K^i\, _j$ denotes the jump of
the extrinsic curvatures $K^i\, _j$ across $\Sigma$.
We will assume that the geometric structure in the region
between the black hole and $\Sigma$ (region I) is described by
the metric
\begin{equation}
ds^2 _1=-(1-\frac{2M_1}{r})dt^2+(1-\frac{2M_1}{r})
^{-1}dr^2+r^2d\Omega^2 _2
\end{equation}
where $d\Omega^2_2$ denotes an infinitesimal interval on
the unit sphere. The geometric structure between $\Sigma$ and
space-like infinity (region II) is taken to be
\begin{equation}
ds^2_2=-(1-\frac{2M_2}{R})dT^2+(1-\frac{2M_2}{R})^{-1}dR^2+\alpha^2R^2d\Omega^2
_2\, .
\end{equation}
In the expression
for $ds^2_2$ the angular part of the metric is enhanced by
a multiplication by a constant $\alpha$ which
is always assumed less
that unity. The extrinsic curvatures
will be computed relative to a unit
space-like vector normal to $\Sigma$ and
one which points in the direction of
increasing radial coordinate.

Let $\Sigma$ be positioned at a
fixed radial distance from the black hole event horizon.
Seen from region I we have $\vec{N}^{(1)}=(g_{rr})^{-1}\partial_r$
and from region II $\vec{N}^{(2)}=(g_{RR})^{-1}\partial_R$.
The fixed position of the wall outside the event horizon
is at $r=r_0$ and $R=R_0$ relative to
the inner and outer geometries respectively. We also
demand that the time-like and the angular coordinates
coincide on the domain-wall. From the continuity of the
gravitational potentials
across $\Sigma$ (except the radial one) we have
\begin{eqnarray}
R_0&=&\frac{1}{\alpha}r_0\\
2r_0(\alpha M_2-M_1)&=&0\, .
\end{eqnarray}
It then follows that
\begin{eqnarray}
\gamma^t\, _t&=&-\frac{1}{r_0^2}(1-\frac{2M_1}{r_0})^{1/2}
(\alpha^2M_2-M_1)\\
\gamma^\theta\, _\theta &=&\frac{(1-\alpha )}{r_0}(1-
\frac{2M_1}{r_0})^{1/2}=
\gamma^\phi\, _\phi\, .
\end{eqnarray}
{}From these expressions we see that the angular components
of the extrinsic curvatures will vanish when $\alpha =1$.
This means that the pressures in the angular directions
in $\Sigma$
also will vanish identically in this limit. In this
limit only the energy density will be non-vanishing
when $M_1\neq M_2$. $\Sigma$ can in that case be interpreted
as a wall composed of pressure less dust.
Relative to a physical observer we will assume that the observed
energy density in $\Sigma$ is positive, $S_{\hat{t}\hat{t}}>0$
 ($\gamma_{\theta\theta}>0$), i.e.
\begin{equation}
1-\alpha^2>\frac{2}{r_0}(M_1-\alpha M_2)\, .
\end{equation}
Combined with the relations which stem from the continuity of the
metric above this is easely seen to be generally satisfied
provided $\alpha^2<1$.
Since $\Sigma$ is assumed to be a domain-wall it follows
that we must have
$S^{\hat{t}}\, _{\hat{t}}=S^{\hat{\theta}}\,
_{\hat{\theta}}=S^{\hat{\phi}}\, _{\hat{\phi}}=\sigma =\mbox{const.} <0$.
This implies in particular that $\gamma^t\, _t=\gamma^\theta\, _\theta
=\gamma^\phi\, _\phi$.

We define in general the gravitational mass $M_G$ measure along the
following lines.
Let $\xi^a$ be a time translation Killing vector
field which is
time-like near infinity such that
$\xi^a\xi_a=-1$
and which has vanishing
norm on the event-horizon. Let $\vec{N}$ be a second Killing
vector field
orthogonal to the event-horizon with normalization such
that $N^aN_a=1$ near infinity.
Let $S$ denote the region outside the
event-horizon of the black hole. The boundary $\partial S$
of $S$ is taken to be the event-horizon $\partial B$
and a two-surface $\partial S_\infty$
at infinity. It then follows that a natural measure on the
gravitational
mass $M_G$ inside $\partial S_\infty$
as measured by a static observer at infinity can
be deduced from \cite{Bardeen,Jensen3}
\begin{eqnarray}
\int_{\partial B+\partial S^{\infty}}\xi^{a;b}d\Sigma_{ab}=
-\int_{S}R^a\, _b \xi^b d\Sigma_a \, .
\end{eqnarray}
$R^a\, _b$ denotes the Ricci tensor.
The integral over $\partial B$ is the surface
gravity $\kappa =N_b\xi^a\nabla_a\xi^b$ multiplied with the
surface area $A$ of $\partial B$.
By the use of this measure it follows that
the total gravitating
mass $M_G$ of the combined black hole domain wall
system relative
to a time-like Killing observer $\vec{\xi}=\partial_T$ at infinity
is given by
$M_G =M_\Sigma +M_H=\alpha^2M_2=\alpha M_1$ where $M_\Sigma$ is the
total gravitating
 mass of $\Sigma$ and $M_H$ is the mass of the black hole
\cite{Jensen3}. We have
specifically that
\begin{eqnarray}
M_\Sigma&=&-8\pi\gamma^\theta\, _{\theta}\\
M_H&=&M_1\, .
\end{eqnarray}
Hence, $M_G<M_H$. The mass of the domain wall is negative due to the large
negative pressures in the angular directions. From these
expressions we can obtain an expression for $\alpha$ in terms
of the mass of the black hole and the mass of $\Sigma$
\begin{equation}
\alpha =\frac{M_1-|M_\Sigma |}{M_1}\, .
\end{equation}
We note that $\alpha$ vanishes when $M_H=|M_\Sigma |$.
This implies that the exterior region outside of $\Sigma$
becomes singular and we will correspondingly confine our
attention to the case when the mass of the hole
is less than the absolute value of the mass of the domain
wall.

We now derive the equation of motion
for a shell collapsing into a black hole positioned at the
center of the wall. Earlier in this investigation the
condition imposed on $\Sigma$ in order for this structure
to describe a domain wall was performed relative to a system
at rest relative to the wall. In the same vein we therefore
introduce Gaussian normal coordinates near every point on $\Sigma$.
The timelike coordinate in this system will be the proper time
$\tau$ of
an observer on the shell while the angular coordinates are
inherited from the surrounding space. Hence on $\Sigma$ we have
coordinates $(\tau ,\theta ,\phi )$. In addition we need a
radial coordinate $\chi$ which measures the proper length from
the observer on the wall and to a point in the vicinity
of $\Sigma$.
Let $\vec{N}$ denote
the unit space-like vector orthogonal to
the collapsing shell $\Sigma$ and $\vec{U}$ the four-velocity
of a mass element of this surface. The orthogonality condition
translates into $\vec{N}\cdot\vec{U}=0$. The four-velocity
in region I relative to the coordinatization eq.(6) is taken as
$\vec{U}=\dot{t}\partial_t+\dot{r}\partial_r$
where $\cdot$ denotes differentiation with respect to
the affine parameter $\tau$. From the
orthogonality condition it then follows that
$\vec{N}=(|g_{tt}|)^{-1}\dot{r}\partial_t+|g_{tt}|\dot{t}\partial_r$
such that $\vec{N}\cdot\vec{N}=+1$ when we assume that $\dot{t}>0$.
The normalization of the four-velocity, $\vec{U}\cdot\vec{U}=-1$, implies
\begin{equation}
g_{tt}\dot{t}=\pm\sqrt{|g_{tt}|+\dot{r}^2}\, .
\end{equation}

The extrinsic curvatures relative to the Gaussian normal
coordinates are simply $K_{\tau\tau} = N_{\tau ;\tau}=U^\mu
U^\nu N_{\mu\nu}$ and
$K_{ij}=N_{i;j}$ ($i,j=\phi\, ,\theta$). The extrinsic curvature
tensor is diagonal such that the $\tau\tau$ component does not
couple to the angular components. However, they are related since
energy-momentum is conserved. In deriving the equation of motion
it is enough to consider the angular components of Israels
equations and we only need to compute the angular components of
the extrinsic curvature tensor. The $\tau\tau$ equation
will express the nature of the forces acting on the wall.
These are not of interest in our investigation.
The gravitational potentials are assumed continuous across
the moving wall. Relative to region I we then have
\begin{eqnarray}
K_{\theta\theta}&=&\frac{1}{2}|g_{tt}|g_{\theta\theta ,r}\dot{t}\\
K_{\phi\phi}&=&\sin^2\theta K_{\theta\theta}\, .
\end{eqnarray}
Similar expressions are derived in region II expressed in terms
of $T$ and $R$. The jump $\gamma^i\, _j$ in the extrinsic
curvatures defined by
$\gamma^i\, _j=K^i\, _j(II)-K^i\, _j(I)$ is
\begin{eqnarray}
\gamma^\theta\, _\theta&=&\gamma^\phi\, _\phi =\frac{(\alpha -1)}{r}
|g_{tt}|\dot{t}\\
\mbox{Tr}\gamma_{ij}&=&3\gamma^\theta\, _\theta
\end{eqnarray}
where the last equility follows since $\Sigma$ models a domain wall.
The $\theta\theta$-equation combined with eq.(17) then leads to
\begin{equation}
\pm\sqrt{|g_{tt}|+\dot{r}^2}=\kappa r
\end{equation}
where we have defined
\begin{equation}
\kappa \equiv\frac{4\pi G\sigma}{(\alpha -1)}>0\, .
\end{equation}
Equation (22) will be our principal equation of motion.
It is useful to define a new radial coordinate
by $z\equiv \kappa r$. We also
define $E\equiv -\kappa^2$ and
\begin{equation}
V(z)=-\kappa^2(z^2+\frac{2M_1\kappa}{z})\, .
\end{equation}
The squared of our principal equation of motion then takes the form
\begin{equation}
(\frac{dz}{d\tau})^2+V(z)=E\, .
\end{equation}
Let us first consider the case when there is no black hole
present. The potential then reduces to $V=-\kappa^2z^2$. Since
$E$ is always negative and non-zero it follows that $\Sigma$ always will
collapse to a certain minimum
 radius and then expand. We can understand
this behaviour as arising due to repulsive self-forces generated by
the large negative pressures.
The behaviour of the collapsing shell is somewhat more complicated
when a black hole is present. The position of the extremum of the potential
function is at
\begin{equation}
z_{\mbox{extr}}=(\frac{\kappa M_1}{\alpha -1})^{1/3}
\end{equation}
where the potential attains its maximal value
\begin{equation}
V_{\mbox{extr}}=V(z_{\mbox{extr}})=-\frac{3M_1\kappa^3}{(M_1\kappa)^{1/3}}\, .
\end{equation}
{}From these equations it is clear that as long as $z_{\mbox{extr}}$
is positioned behind the event horizon of the original
black hole the domain wall will collapse into the
singularity of the hole. It is also clear that there
always exist a class of domain walls for which this holds
irrespective of the black hole parameters. An
explicit computation reveals that this also holds true
in the case of a Reisner-Nordstr\"{o}m black hole.

\section{The cosmic censor and the second law}

We have seen that provided the black hole mass is
large enough compared
with the mass of the innfalling shell it will be
absorbed by the hole and
become part of its singularity structure. From our usual
experience with black holes which absorbs matter with
positive gravitating mass we know that both the pull on
test-particles in the vicinity of the event-horizon
increases as well as the surface-area of the event-horizon.
When a black hole absorbs objects with negative mass
we would naively expect that just the opposite processes
would take place. Indeed, we would expect that the
surface area of the event-horizon where reduced since this
quantity is proportional to $(M_G)^2$ in the usual
Schwartzshild solution. The reduced mass is also similarly
expected to reduce the invariant acceleration felt
by a static observer just outside the horizon.
Also, consider
the
Hawking radiation process. In this process we assume that the
 hole
radiates positive energy to infinity which imply that after a
sufficient
amount of time the amount of the
original matter ``content'' of the singularity which
 contributed
positively to $M_G$
 approaches equality to the negative part of the
singularity which supports the deficit cone.
Hence in a certain limit we would expect that the
gravitating masses of the original singularity pluss the
mass of the structure which supports the deficit cone
would add up to zero. The
singularity
would then display an effective {\it vanishing} gravitating mass which
implies that
the temperature of the black hole will diverge to pluss infinity.
In this limit enough energy will be aviable to excite the
monopole or skyrmion field such that an anti-particle will
be created and hence annihilate with the monopole
(skyrmion) core of the black hole.

These considerations can be tested in a rather
 straightforward way.
Let us first consider the change of the entropy during a
capture process.
Before the hole captures the negative mass carrying object
the entropy is
$S=4\pi M_1^2$ and after the capture
 $S\equiv S_\alpha=4\pi\alpha^{-2}(M_G)^2$ \cite{Jensen3}. From the
matching conditions of the metric
coefficients it follows that $M_G =
\alpha^2M_2=\alpha M_1$
which imply that $dS\equiv S_\alpha -S =0$
\footnote{This result is in sharp contrast to the
result
announced in \cite{Yu} where it was argued that the
entropy
of
the black hole would increase due to the capture
process.}.
The temperature
of the hole is $T\sim M_2^{-1}=\alpha M_1^{-1}<
M_1^{-1}$. Hence, when the black hole swallows the
object its
temperature apparently
decreases, i.e. $dT<0$. The consequences of the
Hawking process
can also
be found quite easely along similar lines. We have that $T\sim
M_2^{-1}=\alpha^2
(M_G )^{-1}=\alpha^2 (M_H-|M_\Sigma |)^{-1}$ ($M_H>|M_
\Sigma |$).
In the Hawking process we assume that $dM_H<0$ while $M_\Sigma$
and $\alpha$ remains
approximately constant for a sufficiently large
hole. It then follows trivially that
 $dT>0$ and $dS<0$ which
conforms with the usual result. It is interesting to
observe that if the possibility $M_H<|M_\Sigma|$
where realizable we will
have a negative $M_G$ which would give rise to a negative
temperature. If we allow $\alpha$ also to be
time dependent in the
radiation process then $T$ should be written as
$T\sim (M_H-|M_\Sigma |)$ which means that the temperatur
would drop to zero when the two mass terms coincide.
However it is
probably not correct to reason along these lines since the relation
eq.(16) is derived on the assumtion that the wall is static.

\section{Conclusion and final remarks}

In this work we have studied a {\it gedanken} experiment
constructed specifically
in order to violate the cosmic censorship hypothesis
and the second law of black hole thermo-dynamics. A near extremal
Reisner-Nordstr\"{o}m black
hole is at the brink of exposing a naked singularity. Either by adding more
charge of some kind \cite{Becken} or some object with negative mass one would
naively expect the cosmic censorship to be broken. In \cite{Becken}
it was shown that adding more charge probably will not make the hole
expose a singularity. In this work we have shown that the area
of the event horizon of a Schwartzshild black hole remains the same when
the hole captures a negative mass object
with a conic deficit angle. This result
can be carried directly over to
the Reisner-Nordstr\"{o}m black hole. Hence, the cosmic censor-ship
hypothesis is not violated when a near extremal black hole captures negative
mass objects of the kind considered in this paper. It also follows
that the second law of thermo-dynamics is not violated either. The effect
of the capture process was apparently
to cool down the black hole while the
entropy of the black hole remained constant, i.e. $dT<0$, $dS=0$.
These considerations does not prove that the censor-ship hypothesis
and the second law holds in any process where a negative mass object
is captured by a black hole but we have managed to reveal a new
mechanism which must be taken into account when
considering such processes. Our treatment has so far been entirely
classical. The possibility is open for the occurrence of new
processes at the quantum level which may modify our results
somewhat. In particular the decrease of the temperature of the black hole
will in a quantum treatment probably find its explanation
in the fact that the assymptotic region is curved.
Assume that the negative mass object is initially sufficiently far
from the black hole such that the line-element eq.(3) describes
the geometry also at a large distance from the black hole with
$\alpha =1$. In such a situation one would second
quantize in the asymptotic region
with respect to the Killing vector $\vec{\xi} =\partial_t$.
After the infall of the object this vector is transformed
into $\vec{\eta}=\alpha\partial_t$. Hence, for a positive
frequency mode $u_k(x)$, i.e. such that ${\cal L}_{\vec{\xi}}
u_k=i\omega u_k$ ($\omega >0$), we have ${\cal L}_{\vec{\eta}}u_k=
\alpha {\cal L}_{\vec{\xi}}u_k$ which implies an effective red-shift
of the modes in the ``out''-vacuum compared to the
modes in the``in''-vacuum. This means that the
vacuum energy is lowered relative to the initial state of the vacuum
something which implies a production of particles
out of the quantum vacuum \footnote{In \cite{Hiscock2,Lousto} the
production of scalar particles from the
associated vacuum due to the formation of
a single global monopole was computed.}. Note that the frequencies
are related by a hole multiplum of $\alpha$ the very same
relation that exist between the temperature of the hole before
and after the innfall of the negative mass object. We plan
to come back to this issue somewhere else.

\section{Acknowledgments}

I thank T.Roman for directing my attention to refs. \cite{Ford2,Ford3}.


\end{document}